\documentclass[a4paper]{article}
\usepackage[ansinew]{inputenc}
\usepackage{times}
\usepackage[T1]{fontenc}
\usepackage{graphicx}
\usepackage{geometry}
\usepackage{amssymb}
\usepackage{amsmath}

\usepackage{rotating}

\usepackage{xcolor}
\colorlet{shadecolor}{gray!25}
\usepackage{subfig}
\usepackage{float}
\usepackage{lscape}

\author{Ludwig A. Hothorn,\\ 
Im Grund 12, D-31867 Lauenau, Germany\\ \scriptsize(retired from Leibniz University Hannover)}
\title{A statistical method for estimating the no-observed-adverse-event-level}
\begin{document}

\maketitle
\begin{abstract}
In toxicological risk assessment the benchmark dose (BMD) is recommended instead of the no-observed-adverse effect-level (NOAEL). Still a simple test procedure to estimate NOAEL is proposed here, explaining its advantages and disadvantages. Versatile applicability is illustrated using four different data examples of selected in vivo toxicity bioassays. 
\end{abstract}

\section{The Problem}

At first glance, it is surprising why BMDL \cite{Jensen2019} is not used as a standard approach today, mainly because of the disadvantages of the NOEAL approach \cite{Kodell2009}. NOAEL depends on the design, i.e. sample size $n_i$, number of doses, chosen dose intervals, kind of endpoints, available historical controls, GLP conditions.  In biomedical research, there are few better standardized studies than the in vivo and in vitro assays in regulatory toxicology, especially with respect to the criteria of minimum sample sizes, recommended number of doses, and a-priori defined endpoints. Thus, in the scenario considered here, the NOAEL does not combine all the described disadvantages of significance testing in mostly unplanned observational studies. Given the many alternative uncertainties, it should be accepted that the NOAEL is 'only' an experimental, i.e. designed dose and not, like BMDL, an interpolated estimate in the hundredths range with a spurious accuracy. Especially since an uncertainty factor with a wide definition range is added to NOAEL for the risk assessment. Practically, the BMD concept has at least four main difficulties: i) its model dependence (which can be reduced by model averaging \cite{Ritz2013}, ii) the necessary choice of BMR, which remains an unresolved challenge for multiple, continuous end points with different scales, variances, and relevance ranges, iii) the considerable width of the confidence interval for designs with only a few doses ( e.g., $k=3+1$ design), and iv) the sometimes non-textbook shape of dose-response dependencies (e.g., pronounced plateaus or even downward effects). e.g., the common $k=3+1$ design), and iv) the sometimes non-textbook shape of dose-response dependencies (e.g., pronounced plateaus or even downward effects at high dose(s)). Furthermore, dose selection in the design phase is more difficult to obtain a smooth log-logistic curve, required for the BMD concept- compared to a design with at least one non-significant and one significant dose (required for the NOAEL concept).\\
 The main difference between NOEAL and BMD lies in the sparseness of assumptions: i) NOEAL assumes only significance of a point null hypothesis and monotonicity of the dose-response relationship for $D_i>NOEAL$ (more precisely the monotonicity of p-values $<0.05$), ii) BMD assumes a perfect fit of a specific nonlinear model as well as interpolation to the dose scale based on a selected BMR value (inverse regression).\\

NOAEL should be in fact estimated as the maximum safe dose, i.e., by a proof-of-safety approach \cite{Hothorn2008} as  step-up non-inferiority tests. This approach is not currently used at all, mainly because it depends on the a priori definition of a threshold value that can still be interpreted, which is difficult practically.  Therefore the simple alternative of a proof-of-hazard approach is used: $NOAEL=MED-1$, where minimum effective dose (MED) is the lowest significant dose, although all higher doses must also be significant.  The Dunnett method \cite{DUNNETT1955} is often used for this purpose.\\
As basic law in toxicological risk assessment, the Paracelsus theorem can be regarded \textit{'All things are poison, and nothing is without poison. The dose alone makes a thing not poison'}. This requires non-significant effects both below and up to the NOAEL, and also significant effects beyond the NOAEL. Based on this model one can derive an order constraint assumption. The Williams procedure \cite{Williams1971} is based exactly on this assumption, but makes no statement about the individual doses (except $D_{max}$).\\
Two approaches to NOAEL estimation are described below, which were derived as special cases of the close testing procedure (CTP) \cite{MARCUS1976} for comparisons to the control and monotonicity assumption.

\section{Closed testing procedures assuming order restriction}
CTP is a general and flexible test principle for multiple testing, which is based on a decision tree of closed partition hypotheses tested by intersection-union-tests (IUT). The definition of the elementary hypotheses of interest is essential. For NOAEL estimation exactly the individual comparisons against control are of interest only, for example  $H_0^{01}=\mu_1-\mu_0$,  $H_0^{02}=\mu_2-\mu_0$,  $H_0^{03}=\mu_3-\mu_0$. Based on these $H_0^\xi$, all intersection hypotheses are created up to the final global hypothesis in order to obtain a closed intersection hypotheses system:\\ 
\textit{partition hypotheses:} $H_0^{012}=\mu_2=\mu_1-\mu_0$, $H_0^{013}=\mu_3=\mu_1-\mu_0$, $H_0^{023}=\mu_3=\mu_2-\mu_0$,\\
 \textit{global hyothesis:} $H_0^{0122}=\mu_3=\mu_2=\mu_1-\mu_0$\\
An elementary hypothesis of interest is rejected if it is rejected exactly, as are all partition hypotheses and the global hypothesis containing them, each with a test at the $\alpha$ level.  Because any level-$\alpha$-test can be used, the closure test is flexible. If one assumes one-sided tests with an order restriction $\mu_0\leq \mu_1\leq...\leq \mu_k$ (for any pattern of equalities and inequalities, but at least $\mu_0< \mu_k$)  due to a monotonic dose-response relationship, the following applies: if $H_0^{0122}=\mu_3=\mu_2=\mu_1-\mu_0$ is rejected, $H_0^{013}=\mu_3=\mu_1-\mu_0$, $H_0^{023}=\mu_3=\mu_2-\mu_0$, $H_0^{03}=\mu_3-\mu_0$ must also be rejected, without testing it. This allows the use of simple pairwise tests for \textit{all} hypotheses. Here, pairwise contrast tests are used \cite{Hothorn2021} to exploit the full degree of freedom of the k-sampling problem.\\
The adjusted p-value for the elementary hypothesis of interest results from the maximum of all p-values in its hypothesis tree (IUT):\\
$H_0^{01}: H_0^{01} \subset H_0^{012} \subset H_0^{0123} \Rightarrow p_{01}=max(p_{01}, p_{02}, p_{03})$\\
$H_0^{02}: H_0^{02} \subset H_0^{012} \subset H_0^{0123} \Rightarrow p_{02}=max(p_{02}, p_{03})$\\
$H_0^{03}: H_0^{03} \subset H_0^{0123} \Rightarrow p_{03}= p_{03}$\\

From a toxicological point of view, it is interesting that the comparison $D_{max}-0$ is performed with an unadjusted level $\alpha$-test (and thus of maximum possible power). The $D_{max-1}-0, D_{max-2}-0,...$ comparisons with slightly higher, increasing conservatism (because on the IUT condition). This is an intuitive, simple and powerful method for NOAEL estimation. In particular, it can be generalized to many situations that occur in toxicology: parametric and non-parametric tests, assuming homogeneous variance or not, considering proportions, counts, poly-k-estimates, using a mixed effect model (e.g. to take within/between-litter dependencies into account), etc.. Related R-software is available, see examples in the Appendix.

\section{Demonstration the procedures for four selected in-vivo bioassay}
The first example contains body weight changes for Fischer 344 rats treated with 0, 100, 200, 500, and 750 mg/kg doses of aconiazide over a 14-day period \cite{West2005}. The second example considers relative kidney weight data in a feeding study with a crop-protecting compound \cite{Tamhane2004}. The third example use severity scores of lesions in epithelium after exposure to formaldehyde in doses of 0, 2, 6, and 15 ppm in Fisher 344 rats \cite{Yanagawa1997}. The fourth example considers effect of vinylcyclohexene diepoxide exposure to 0 mg/ml, (group 0), 25 mg/ml (group 1), 50 mg/ml (group 2), and 100 (group 3) mg/ml, on the incidence of alveolar/bronchiolar tumors in a long term rodent carcinogenicity study on female B6C3F1 mice \cite{Piegorsch1997}. Figure 1 shows three boxplots (for examples 1,2,3 from the left to the right, and the mosaicplot for the crude tumor proportions in example 4, right panel). This random selection shows designs with only 3 dose groups (and 4 dose groups in example 1).  

\begin{figure}[htbp]
	\centering
		\includegraphics[width=0.232\textwidth]{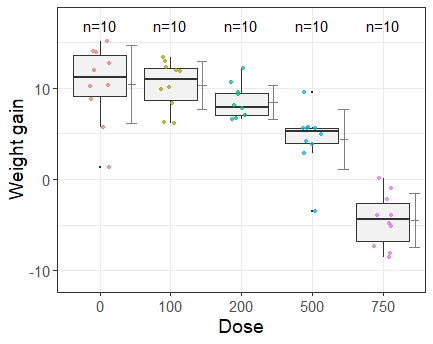}
		\includegraphics[width=0.232\textwidth]{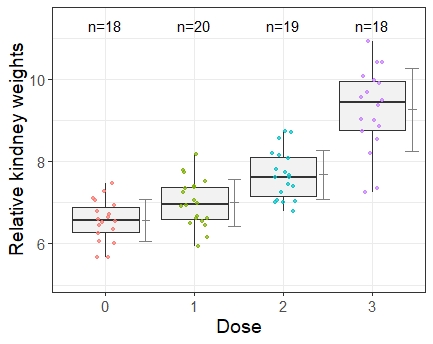}
		\includegraphics[width=0.232\textwidth]{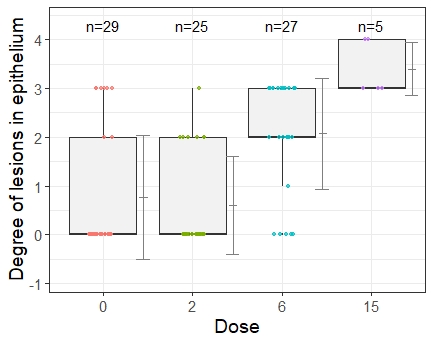}
		\includegraphics[width=0.232\textwidth]{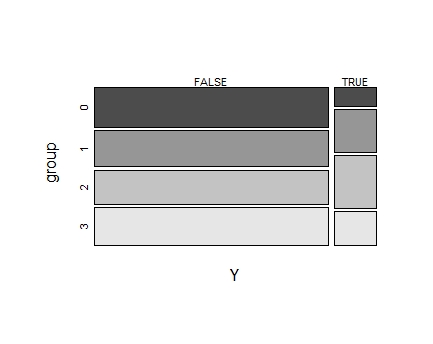}
	\caption{Raw data plots for the four data examples (from left to right examples 1, 2, 3, 4)}
	\label{fig:BoxTamh}
\end{figure}
Since normal distribution and homogeneous variances can be assumed for the endpoint body weight gain in example 1, the closure test with parametric multiple contrast is estimated. (R-Code see Appendix 1). In the original paper of example 2 ratio-to-control tests under variance heterogeneity were used \cite{Tamhane2004}. Therefore a closure test is used by simultaneous ratio-to-control tests \cite{Hothorn2011} (R-Code see Appendix 2). The severity score endpoint in example 3 is analysed by a nonparametric multiple contrast test for the relative effect size \cite{Konietschke2012} (p-value in brackets) and alternatively by contrast test using the most likely transformation model for count data \cite{Siegfried2020} (R-Code see Appendix 3). In long-term carcinogenicity bioassay, the complex relationship between tumor development and mortality can be modeled with the poly-k-test \cite{BIELER1993} whereas related multiple contrast tests are available in the R package MCPAN \cite{Schaarschmidt2008}. This uses a generalized contrast test for the difference of proportions, more precisely, weighted proportions. (R-Code see Appendix 4). Table 1 shows for these 4 examples the adjusted p-values of the specific closure tests for the respective control comparisons and the estimated NOAEL (marked in bold). In example 2, no NOAEL can be estimated as there is no non-significant dose, contrary to the original work \cite{Tamhane2004}, so the control 0 is marked in bold (an inappropriate dose selection in this assay). The nonparametric approach in example 3 reveals the same decision as the most likely transformation approach, but with slightly different p-values $(0.058; 0.00061; <0.00001)$.

\begin{table}[ht]
\centering
\begin{tabular}{ll|ll|ll|ll}
\multicolumn{2}{c}{Exa. 1} & \multicolumn{2}{c}{Exa. 2} & \multicolumn{2}{c}{Exa. 3} & \multicolumn{2}{c}{Exa. 4} \\
 Comp.    &  $p^{adj}$    &    Comp.    & $p^{adj}$    	& Comp.     & $p^{adj}$          & Comp.         & $p^{adj}$          \\ \hline

 100-0    &  0.48    		&    1/\textbf{0}    & 0.012   		&  \textbf{2}-0   & 0.058  	&  \textbf{25}-0      &    0.059                \\
 \textbf{200}-0 &  0.11 &    2/0    & $<0.00001$   &    6-0       &   0.00016    					&  50-0         &  0.017        \\
 500-0    &  0.00086  	&    3/0   & $<0.00001$			&    15-0       &   $<0.00001$ 				& 100-0        &  0.017         \\
 750-0    &  $<0.00001$ &    -   & ... 						&     -   			& ...             				& -       					&    ...      \\
   
\end{tabular}
\caption{Adjusted p-values and estimated NOAEL (bold) for the four examples} 
\end{table}
These examples have been selected to demonstrate that this approach can be applied to numerous scenarios, i.e. different endpoints, different effect sizes, different test principles, different assumptions.\\

To assess the difference between without/with assumption of order restriction, the p-values for the Dunnett procedure and the CTP are compared directly in Table 2. The same NOAEL is estimated in both approaches. The p-values differ slightly, but within the accuracy range of a multivariate t-distribution.

\begin{table}[ht]
\centering
\begin{tabular}{ll|ll}
\multicolumn{2}{c}{CTP} & \multicolumn{2}{c}{Dunnett procedure}  \\ \hline \hline
 Comp.    &  $p^{adj}$  &    Comp.    		& $p^{adj}$    	    \\ \hline

 100-0    &  0.483    		&    100-0          & 0.960   		     \\
 \textbf{200}-0 &  0.1108 &   \textbf{200-0}    				& 0.173          \\
 500-0    &  0.000864  	&    500-0   					& $9.01e-05$			        \\
 750-0    &  $1.969225e-11$ &    750-0 						&   $ 5.37e-14$     \\
   
\end{tabular}
\caption{Adjusted p-values for Dunnett procedure  and CTP in example 1} 
\end{table}
\section{Conclusions} 
An intuitive, simple and flexible closure test with pairwise contrast tests is proposed to estimate NOAEL. It is based only on the assumption of a monotonic dose-response relationship. Notice, all discussed disadvantages of the NOAEL approach compared to the BMD approach remain. Nevertheless, it can be recommended as a possibility for the usual routine evaluation of in vivo and in vitro bioassays in regulatory toxicology, particularly for the common-used designs with a few doses or concentrations only. Extensions for multiple primary endpoints, e.g. multiple tumors, and the mixed model for litter-dependencies in reprotoxicological bioassays will be described next.

\footnotesize
\subsubsection*{Acknowledgement} 
 Dr. Christian Ritz, University of Copenhagen I thank for valuable suggestions for a better visualization of the approach.





\footnotesize
\begin{thebibliography}{10}
\expandafter\ifx\csname url\endcsname\relax
  \def\url#1{\texttt{#1}}\fi
\expandafter\ifx\csname urlprefix\endcsname\relax\def\urlprefix{URL }\fi
\expandafter\ifx\csname href\endcsname\relax
  \def\href#1#2{#2} \def\path#1{#1}\fi

\bibitem{Kodell2009}
R.~L. Kodell, Replace the noael and loael with the bmdl01 and bmdl10,
  Environmental and Ecological Statistics 16~(1) (2009) 3--12.
\newblock \href {http://dx.doi.org/10.1007/s10651-007-0075-3}
  {\path{doi:10.1007/s10651-007-0075-3}}.

\bibitem{Jensen2019}
S.~M. Jensen, F.~M. Kluxen, C.~Ritz, A review of recent advances in benchmark
  dose methodology, Risk Analysis 39~(10) (2019) 2295--2315.
\newblock \href {http://dx.doi.org/10.1111/risa.13324}
  {\path{doi:10.1111/risa.13324}}.

\bibitem{HH2000}
L.~A. Hothorn, D.~Hauschke, Identifying the maximum safe dose: a multiple testing
  approach,  Biopharmaceutical Statistics 10 (2000) 15-30.

\bibitem{Hothorn2008}
L.~A. Hothorn, M.~Hasler, Proof of hazard and proof of safety in toxicological studies using simultaneous confidence intervals for differences and ratios to control, J of Biopharmaceutical Statistics 18 (2008) 915-933.

\bibitem{Tamhane2001}
A.~C. Tamhane, C.~W. Dunnett, J.~W. Green, J.~D. Wetherington, Multiple test
  procedures for identifying the maximum safe dose, Journal of the American
  Statistical Association 96~(455) (2001) 835--843.
\newblock \href {http://dx.doi.org/10.1198/016214501753208546}
  {\path{doi:10.1198/016214501753208546}}.

\bibitem{DUNNETT1955}
C.~W. Dunnett, A multiple comparison procedure for comparing several treatments
  with a control, Journal of the American Statistical Association 50~(272)
  (1955) 1096--1121.

\bibitem{Williams1971}
D.~Williams, A test for differences between treatment means when several dose
  levels are compared with a zero dose control, Biometrics 27 (1971) 103-117.

\bibitem{MARCUS1976}
R.~Marcus, E.~Peritz, K.~R. Gabriel, Closed testing procedures with special
  reference to ordered analysis of variance, Biometrika 63~(3) (1976) 655--660.
\newblock \href {http://dx.doi.org/10.1093/biomet/63.3.655}
  {\path{doi:10.1093/biomet/63.3.655}}.

\bibitem{Hothorn2016}
L.~A. Hothorn, The two-step approach-a significant ANOVA F-test before
  Dunnett's comparisons against a control-is not recommended, Communications in
  Statistics- Theory and Methods 45~(11) (2016) 3332--3343.
\newblock \href {http://dx.doi.org/10.1080/03610926.2014.902225}
  {\path{doi:10.1080/03610926.2014.902225}}.

\bibitem{Hothorn2021}
L.~A. Hothorn, Closed test procedures for the comparison of dose groups against a negative control group or placebo. arXiv 2011.13758v1

\bibitem{West2005}
R.~W. West, R.~L. Kodell, Changepoint alternatives to the noael, Journal of
  Agricultural Biological and Environmental Statistics 10~(2) (2005) 197--211.
\newblock \href {http://dx.doi.org/10.1198/108571105X46525}
  {\path{doi:10.1198/108571105X46525}}.

\bibitem{Tamhane2004}
A.~Tamhane, B.~Logan, Finding the maximum safe dose level for heteroscedastic
  data, Journal of Biopharmaceutical Statistics 14 (2004) 843-856.

\bibitem{Yanagawa1997}
T.~Yanagawa, Y.~Kikuchi, K.~G. Brown, No-observed-adverse-effect levels in
  severity data, Journal of the American Statistical Association 92~(438)
  (1997) 449--454.

\bibitem{Piegorsch1997}
W.W.~Piegorsch, A.J.~Bailer (1997): Statistics for environmental biology and
  toxicology. Chapman and Hall, London. Table 6.5, page 238.

\bibitem{Hothorn2011}
L.~A. Hothorn, G.~D. Djira, A ratio-to-control Williams-type test for trend,
  Pharmaceutical Statistics 10~(4) (2011) 289--292.
\newblock \href {http://dx.doi.org/10.1002/pst.464}
  {\path{doi:10.1002/pst.464}}.

\bibitem{Konietschke2012}
F.~Konietschke, L.~A. Hothorn, Rank-based multiple test procedures and
  simultaneous confidence intervals, Electronic Journal of Statistics 6 (2012)
  738--759.
\newblock \href {http://dx.doi.org/10.1214/12-EJS691}
  {\path{doi:10.1214/12-EJS691}}.


\bibitem{Ritz2013}
C.~Ritz, D.~Gerhard, L.~A. Hothorn, A unified framework for benchmark dose estimation applied to mixed models and model averaging, Statistics in Biopharm. Res. 5 (2013)
  79--90.

\bibitem{Siegfried2020}
S.~Siegfried, T.~Hothorn, Count transformation models, Methods in Ecology and
  Evolution 11~(7) (2020) 818--827.
\newblock \href {http://dx.doi.org/10.1111/2041-210X.13383}
  {\path{doi:10.1111/2041-210X.13383}}.

\bibitem{BIELER1993}
G.~S. Bieler, R.~L. Williams, Ratio estimates, the delta method, and quantal
  response tests for increased carcinogenicity, Biometrics 49~(3) (1993)
  793--801.
\newblock \href {http://dx.doi.org/10.2307/2532200}
  {\path{doi:10.2307/2532200}}.

\bibitem{Schaarschmidt2008}
F.~Schaarschmidt, M.~Sill, L.~A. Hothorn, Approximate simultaneous confidence
  intervals for multiple contrasts of binomial proportions, Biometrical Journal
  50~(5) (2008) 782--792.
\newblock \href {http://dx.doi.org/10.1002/bimj.200710465}
  {\path{doi:10.1002/bimj.200710465}}.

\end{thebibliography}
\section*{Appendix}
\footnotesize
\subsection*{Example 1}
\scriptsize
\begin{verbatim}
Wes <-
structure(list(Gain = c(5.7, 10.2, 13.9, 10.3, 1.3, 12, 14, 15.1, 
8.8, 12.7, 8.3, 12.3, 6.1, 10.1, 6.3, 12, 13, 13.4, 11.9, 9.9, 
9.5, 8.1, 7, 7.8, 9.3, 12.2, 6.7, 10.6, 6.6, 7, 2.9, 5.6, -3.5, 
9.5, 5.7, 4.9, 3.8, 5.6, 5.6, 4.2, -8.6, 0.1, -3.9, -4, -7.3, 
-2.2, -5.2, -1, -8.1, -4.8), dose = c(0L, 0L, 0L, 0L, 0L, 0L, 
0L, 0L, 0L, 0L, 100L, 100L, 100L, 100L, 100L, 100L, 100L, 100L, 
100L, 100L, 200L, 200L, 200L, 200L, 200L, 200L, 200L, 200L, 200L, 
200L, 500L, 500L, 500L, 500L, 500L, 500L, 500L, 500L, 500L, 500L, 
750L, 750L, 750L, 750L, 750L, 750L, 750L, 750L, 750L, 750L), 
    Dose = structure(c(1L, 1L, 1L, 1L, 1L, 1L, 1L, 1L, 1L, 1L, 
    2L, 2L, 2L, 2L, 2L, 2L, 2L, 2L, 2L, 2L, 3L, 3L, 3L, 3L, 3L, 
    3L, 3L, 3L, 3L, 3L, 4L, 4L, 4L, 4L, 4L, 4L, 4L, 4L, 4L, 4L, 
    5L, 5L, 5L, 5L, 5L, 5L, 5L, 5L, 5L, 5L), .Label = c("0", 
    "100", "200", "500", "750"), class = "factor")), row.names = c(NA, 
-50L), class = "data.frame")
ni<-aggregate(Gain ~ Dose, data = Wes, length)$Gain # sample sizes
library(multcomp)
library(sandwich)
mod1<-lm(Gain~Dose, data=Wes)
CM04 <- c(-1,0,0,0,1)
CM03 <- c(-1,0,0,1,0)
CM02 <- c(-1,0,1,0,0)
CM01 <- c(-1,1,0,0,0) # pairwise contrasts

cmat0123<-contrMat(ni[1:4], type="Williams"); V4 <-c(0,0,0)
Cmat0123<-cbind(cmat0123,V4)
cmat012<-contrMat(ni[1:3], type="Williams"); V3 <-c(0,0)
Cmat012<-cbind(cmat012,V3,V3) # subset Williams-type contrasts

DuS<-summary(glht(mod1, linfct = mcp(Dose ="Dunnett"), vcov =vcovHC, alternative="less"))$test$pvalues
T04<-summary(glht(mod1, linfct = mcp(Dose = CM04),vcov =vcovHC,alternative="less"))$test$pvalues
T03<-summary(glht(mod1, linfct = mcp(Dose = CM03),vcov =vcovHC,alternative="less"))$test$pvalues
T02<-summary(glht(mod1, linfct = mcp(Dose = CM02),vcov =vcovHC,alternative="less"))$test$pvalues
T01<-summary(glht(mod1, linfct = mcp(Dose = CM01),vcov =vcovHC,alternative="less"))$test$pvalues

W01234<-summary(glht(mod1, linfct = mcp(Dose ="Williams"),vcov =vcovHC,alternative="less"))$test$pvalues
W0123<-summary(glht(mod1, linfct = mcp(Dose =Cmat0123),vcov =vcovHC,alternative="less"))$test$pvalues
W012<-summary(glht(mod1, linfct = mcp(Dose =Cmat012),vcov =vcovHC,alternative="less"))$test$pvalues
W01<-T01

CTP4<-T04
CTP3<-max(T03,T04)
CTP2<-max(T04,T03,T02)
CTP1<-max(T04,T03,T02, T01) # closure test using pairwise contrasts

CTW4<-W01234
CTW3<-max(W01234,W0123)
CTW2<-max(W01234,W0123,W012)
CTW1<-max(W01234,W0123, W012, W01) # closure test using subset Williams contrasts- not discussed in the paper

\end{verbatim}
\footnotesize
\subsection*{Example 2}
\scriptsize
\begin{verbatim}
Tamh <-
        structure(list(dose = c(0L, 0L, 0L, 0L, 0L, 0L, 0L, 0L, 0L, 0L,
                                0L, 0L, 0L, 0L, 0L, 0L, 0L, 0L, 1L, 1L, 1L, 1L, 1L, 1L, 1L, 1L,
                                1L, 1L, 1L, 1L, 1L, 1L, 1L, 1L, 1L, 1L, 1L, 1L, 2L, 2L, 2L, 2L,
                                2L, 2L, 2L, 2L, 2L, 2L, 2L, 2L, 2L, 2L, 2L, 2L, 2L, 2L, 2L, 3L,
                                3L, 3L, 3L, 3L, 3L, 3L, 3L, 3L, 3L, 3L, 3L, 3L, 3L, 3L, 3L, 3L,
                                3L), 
																kidneywt = c(6.593, 7.48, 6.93, 5.662, 6.789, 7.268, 6.647,
                                6.443, 6.713, 6.057, 6.253, 7.045, 6.552, 5.668, 6.354, 6.511,
                                7.111, 6.015, 7.062, 7.347, 7.733, 7.396, 8.173, 6.938, 6.988,
                                6.621, 7.508, 6.657, 7.787, 6.537, 7.369, 6.623, 6.456, 6.507,
                                6.154, 5.934, 6.909, 7.252, 7.006, 8.706, 7.257, 7.743, 7.026,
                                8.561, 7.674, 7.45, 8.188, 8.15, 7.619, 8.722, 7.387, 6.798,
                                7.617, 8.071, 7.02, 7.821, 7.063, 9.569, 9.362, 10.911, 9.961,
                                9.497, 9.911, 8.544, 10.404, 10.421, 10.065, 9.67, 8.194, 8.989,
                                7.347, 7.26, 9.017, 8.847, 8.723),
        Dose = structure(c(1L, 1L, 1L, 1L, 1L, 1L, 1L, 1L, 1L, 1L, 1L, 1L, 1L, 1L, 1L, 1L, 1L, 1L, 1L,
                          2L, 2L, 2L, 2L, 2L, 2L, 2L, 2L, 2L, 2L, 2L, 2L, 2L, 2L, 2L, 2L, 2L, 2L, 2L, 2L,
                          3L, 3L, 3L, 3L, 3L, 3L, 3L, 3L, 3L, 3L, 3L, 3L, 3L, 3L, 3L, 3L, 3L, 3L,
                          4L, 4L, 4L, 4L, 4L, 4L, 4L, 4L, 4L, 4L, 4L, 4L, 4L, 4L, 4L, 4L, 4L, 4L),
                                        .Label = c("0", "1", "2",  "3"), class = "factor")),
                                row.names = c(NA, -75L), class = "data.frame")

library(mratios)
 RcmatDu<-contrMatRatio(ni, type="Dunnett")
 cm01numC<-rbind(RcmatDu$numC[1,],RcmatDu$numC[1,])
 cm01denC<-rbind(RcmatDu$denC[1,],RcmatDu$denC[1,])
 cm02numC<-rbind(RcmatDu$numC[2,],RcmatDu$numC[2,])
 cm02denC<-rbind(RcmatDu$denC[2,],RcmatDu$denC[2,])
 cm03numC<-rbind(RcmatDu$numC[3,],RcmatDu$numC[3,])
 cm03denC<-rbind(RcmatDu$denC[3,],RcmatDu$denC[3,]) # pseudo double pairwise contrasts

 R03<-simtest.ratioVH(kidneywt~Dose, data=Tamh,Num.Contrast =  cm03numC,
                        Den.Contrast = cm03denC,alternative="greater")$p.value.adj
 R02<-simtest.ratioVH(kidneywt~Dose, data=Tamh,Num.Contrast = cm02numC,
                      Den.Contrast = cm02denC,alternative="greater")$p.value.adj
 R01<-simtest.ratioVH(kidneywt~Dose, data=Tamh,Num.Contrast =  cm01numC,
                      Den.Contrast = cm01denC,alternative="greater")$p.value.adj
 CTR3<-R03
 CTR2<-max(R03,R02)
 CTR1<-max(R03,R02, R01) # CTP using pairwise subset contrasts

\end{verbatim}

\footnotesize
\subsection*{Example 3}
\scriptsize
\begin{verbatim}
Epi <-
     structure(list(dose = c(0L, 0L, 0L, 0L, 0L, 0L, 0L, 0L, 0L, 0L,
     0L, 0L, 0L, 0L, 0L, 0L, 0L, 0L, 0L, 0L, 0L, 0L, 0L, 0L, 0L, 0L,
     0L, 0L, 0L, 2L, 2L, 2L, 2L, 2L, 2L, 2L, 2L, 2L, 2L, 2L, 2L, 2L,
     2L, 2L, 2L, 2L, 2L, 2L, 2L, 2L, 2L, 2L, 2L, 2L, 6L, 6L, 6L, 6L,
     6L, 6L, 6L, 6L, 6L, 6L, 6L, 6L, 6L, 6L, 6L, 6L, 6L, 6L, 6L, 6L,
     6L, 6L, 6L, 6L, 6L, 6L, 6L, 15L, 15L, 15L, 15L, 15L),
     score = c(0L, 0L, 0L, 0L, 0L, 0L, 0L, 0L, 0L, 0L, 0L, 0L, 0L, 0L, 0L, 0L, 0L,
               0L, 0L, 0L, 0L, 2L, 2L, 3L, 3L, 3L, 3L, 3L, 3L, 0L, 0L, 0L, 0L,
               0L, 0L, 0L, 0L, 0L, 0L, 0L, 0L, 0L, 0L, 0L, 0L, 0L, 0L, 2L, 2L,
               2L, 2L, 2L, 2L, 3L, 0L, 0L, 0L, 0L, 0L, 1L, 2L, 2L, 2L, 2L, 2L,
               2L, 2L, 2L, 3L, 3L, 3L, 3L, 3L, 3L, 3L, 3L, 3L, 3L, 3L, 3L, 3L,
               3L, 3L, 3L, 4L, 4L),
     Dose = structure(c(1L, 1L, 1L, 1L, 1L, 1L,
            1L, 1L, 1L, 1L, 1L, 1L, 1L, 1L, 1L, 1L, 1L, 1L, 1L, 1L, 1L, 1L,
            1L, 1L, 1L, 1L, 1L, 1L, 1L, 2L, 2L, 2L, 2L, 2L, 2L, 2L, 2L, 2L,
            2L, 2L, 2L, 2L, 2L, 2L, 2L, 2L, 2L, 2L, 2L, 2L, 2L, 2L, 2L, 2L,
            3L, 3L, 3L, 3L, 3L, 3L, 3L, 3L, 3L, 3L, 3L, 3L, 3L, 3L, 3L, 3L,
            3L, 3L, 3L, 3L, 3L, 3L, 3L, 3L, 3L, 3L, 3L, 4L, 4L, 4L, 4L, 4L
            ), .Label = c("0", "2", "6", "15"), class = "factor")),
     row.names = c(NA, -86L), class = "data.frame")



########## most likely count data transformation model
ni<-aggregate(score ~ Dose, data = Epi, length)$score
library(cotram)
mod2<-cotram(score~Dose, data=Epi,order=4)
DuC<-summary(glht(mod2, linfct = mcp(Dose = "Dunnett"), alternative="greater"))
CM03 <- c(-1,0,0,1)
CM02 <- c(-1,0,1,0)
CM01 <- c(-1,1,0,0)
CC03<-summary(glht(mod2, linfct = mcp(Dose =CM03), alternative="greater"))$test$pvalues
CC02<-summary(glht(mod2, linfct = mcp(Dose =CM02), alternative="greater"))$test$pvalues
CC01<-summary(glht(mod2, linfct = mcp(Dose =CM01), alternative="greater"))$test$pvalues

CTPC3<-CC03
CTPC2<-max(CC02,CC03)
CTPC1<-max(CC03,CC02, CC01) # closure test using pairwise contrasts

#################### nonparametric multiple pairwise contrasts
library(nparcomp)
CN03 <- rbind(c(-1,0,0,1), c(-1,0,0,1))
CN02 <- rbind(c(-1,0,1,0), c(-1,0,1,0))
CN01 <- rbind(c(-1,1,0,0), c(-1,1,0,0)) # pseudo double pairwise contrasts

DuN<-summary(nparcomp(score~Dose, data=Epi, type="Dunnett", alternative="greater"))
NP03<-nparcomp(score~Dose, data=Epi, type="UserDefined", contrast.matrix=CN03, alternative="greater")$Analysis$p.Value
NP02<-nparcomp(score~Dose, data=Epi, type="UserDefined", contrast.matrix=CN02, alternative="greater")$Analysis$p.Value
NP01<-nparcomp(score~Dose, data=Epi, type="UserDefined", contrast.matrix=CN01, alternative="greater")$Analysis$p.Value

CTPNP3<-NP03
CTPNP2<-max(NP03,NP02)
CTPNP1<-max(NP03,NP02,NP01) # closure test using pairwise contrasts
\end{verbatim}
\footnotesize
\subsection*{Example 4}
\scriptsize
\begin{verbatim}
library(MCPAN)
data(bronch)
ni<-aggregate(Y ~ group, data = bronch, length)$Y
cm03 <- rbind(CM03,CM03)
cm02 <- rbind(CM02,CM02)
cm01 <- rbind(CM01,CM01) #pseudo double pairwise contrasts

D03<-poly3test(time=bronch$time, status=bronch$Y,
     f=bronch$group, cmat =cm03, method = "BW", alternative="greater")$p.val.adj
D02<-poly3test(time=bronch$time, status=bronch$Y,
     f=bronch$group, cmat =cm02, method = "BW", alternative="greater")$p.val.adj
D01<-poly3test(time=bronch$time, status=bronch$Y,
     f=bronch$group, cmat =cm01, method = "BW", alternative="greater")$p.val.adj

CTPC3<-D03
CTPC2<-max(D03,D02)
CTPC1<-max(D03,D02,D01) # closure test using pairwise contrasts
\end{verbatim}

\end{document}